\documentclass[12pt]{article} 
\usepackage[utf8]{inputenc}    
\usepackage{graphicx}
\usepackage{multirow}
\usepackage{amsmath,amssymb,amsfonts}
\usepackage{amsthm}
\usepackage{mathrsfs}
\usepackage[title]{appendix}
\usepackage{xcolor}
\usepackage{textcomp}
\usepackage{manyfoot}
\usepackage{booktabs}
\usepackage{algorithm}
\usepackage{algorithmicx}
\usepackage{algpseudocode}
\usepackage{adjustbox}
\usepackage{listings}
\usepackage[margin=1cm]{caption}
\usepackage{subcaption}
\usepackage{url}
\usepackage{geometry}          
\begin{document}

\title{The Dataset of Daily Air Quality for the Years 2013-2023 in Italy}

\author{
Alessandro Fusta Moro$^{1,*}$, 
Alessandro Fassò$^{2}$ and
Jacopo Rodeschini$^{1}$\\  \footnotesize
$^{1}$Department of Engineering and Applied Science, University of Bergamo, Dalmine, Italy \\ \footnotesize
$^{2}$Department of Economics, University of Bergamo, Bergamo, Country \\ \footnotesize
*Corresponding author: alessandro.fustamoro@unibg.it
}
\date{\today}  

\maketitle

\abstract{Air quality and climate are major issues in Italian society and lie at the intersection of many research fields, including public health and policy planning. There is an increasing need for readily available, easily accessible, ready-to-use and well-documented datasets on air quality and climate. In this paper, we present the GRINS\_\allowbreak AQCLIM dataset, created under the GRINS project framework covering the Italian domain for an extensive time period. It includes daily statistics (e.g., minimum, quartiles, mean, median and maximum) for a collection of air pollutant concentrations and climate variables at the locations of the 700+ available monitoring stations. Input data are retrieved from the European Environmental Agency and Copernicus Programme and were subjected to multiple processing steps to ensure their reliability and quality. These steps include automatic procedures for fixing raw files, manual inspection of stations' information, the detection and removal of anomalies, and the temporal harmonisation on a daily basis. Datasets are hosted on Zenodo under open-access principles.}

\section*{Background \& Summary}
In this paper, we present a new dataset for air quality in Italy called “GRINS\_\allowbreak AQCLIM\_\allowbreak points\_\allowbreak Italy", short “GRINS\_\allowbreak AQCLIM dataset", created under the framework of the Growing Resiliant INclusive and Sustainable (GRINS) project, funded by the NextGeneration EU. 

The GRINS\_\allowbreak AQCLIM dataset is based on the measurements of air pollutant concentrations recorded by the Italian ground-based monitoring network managed by the Regional Agencies for the Environmental Protection (ARPAs). The dataset provides daily summary statistics at 744 locations over the 11-year period from 2013 to 2023. The airborne pollutants considered are: nitrogen monoxide and dioxide (NO and NO$_2$), particulate matter (PM$_{10}$ and PM$_{2.5}$), ozone (O$_3$), ammonia (NH$_3$), carbon monoxide (CO), and sulphur dioxide (SO$_2$), along with various climate variables useful for modelling purpose (i.e., the height of the atmosphere boundary layer, vegetation indexes, relative humidity, temperature, solar radiation, precipitation, wind speed and direction).

\subsection*{Italian air quality and data availability}

The need for a dataset on air quality in Italy is driven by the high levels of air pollution, especially in some areas, such as the Po Valley (see e.g., \cite{eu_quality,otto2024spatiotemporal}). Air quality data have been widely used across numerous scientific contexts. In Italy, these data have been extensively utilised to investigate health impacts, including studies on pregnancy outcomes \cite{capobussi2016air}, COVID-19 mortality (e.g., \cite{coker2020effects, gatti2020machine}), ruptured abdominal aneurysms \cite{bozzani2024spatio}, pulmonary fibrosis \cite{conti2018association}, and neurodegenerative conditions such as dementia and Parkinson's disease \cite{trentalange2025association}, among numerous other applications. Furthermore, air quality data have been used as crucial inputs for scenario analyses concerning green urban planning and city management (e.g., \cite{bottalico2016air, manes2016regulating, sebastiani2021modeling}), as well as for evaluating emission reduction strategies \cite{rodeschini2024scenario}. Air quality data also present a significant opportunity for fair comparison of different spatio-temporal statistical models \cite{otto2024spatiotemporal}. 

\begin{sloppypar}
In recent years, several initiatives have emerged to provide readily accessible air quality datasets for Italy. Notably, the Agrimonia project \cite{fasso2023agrimonia} focused on correlating particulate matter concentrations with livestock activities and published a daily dataset on Zenodo (\url{https://zenodo.org/records/7956006}) covering Lombardy from 2016 to 2021, which has garnered over 6,000 downloads. Concurrently to ground-based monitoring stations, the European Copernicus programme offers hourly air pollutant concentration estimates on a 0.1° x 0.1° grid across Europe, derived from an ensemble of Chemical Transport Models \cite{marecal2015regional}. Nevertheless, these data frequently exhibit limitations in capturing high peaks \cite{CAMS_EQC_IRA_2022}. Within Italy, ISPRA systematically publishes yearly summary statistics, including threshold exceedances and quantiles, for individual monitoring stations, on its official website, encompassing a comprehensive period from 2002 to 2022 (\url{https://www.isprambiente.gov.it/it/banche-dati/banche-dati-folder/aria/qualita-dellaria}). Beyond the European context, readily available datasets for various air pollutants in China are also hosted on Zenodo \cite{wei2021chinahighpm10, wei2023ground}. Addressing the existing fragmented landscape, the proposed GRINS\_\allowbreak AQCLIM aims to standardize access to Italian air quality data initially for the 2013-2023 period, with the prospect of extending both backward and forward in time.

The GRINS\_\allowbreak AQCLIM dataset contains air pollutant levels recorded by the monitoring stations, managed by the regional agencies for environmental protection (ARPAs). Observed data are submitted from the ARPAs to the European Environmental Agency (EEA) through the air quality portal (\url{https://aqportal.discomap.eea.europa.eu}), where air quality data for all of Europe can be accessed. Currently, direct access to Italian air quality data is available via regional ARPAs' websites and the European Environment Agency (EEA) air quality portal. Obtaining comprehensive raw data from regional ARPAs often proves challenging due to incomplete datasets across regions, missing periods, and a lack of API services for automated retrieval. While the EEA offers API access, its raw data frequently necessitate extensive quality control. Common issues encountered within EEA raw files include duplicate station entries across different files, inconsistent temporal resolutions for the same station (e.g., hourly and daily within the same dataset), internal temporal resolution shifts, and overlapping bi-hourly and hourly records. Consequently, data acquired from the EEA need pre-use transformations and validation, as demonstrated in existing researches (e.g., \cite{fioravanti2022spatiotemporal, fasso2023agrimonia}). These essential quality assurance procedures often involve statistical steps, such as anomaly detection and removal, or missing-data imputation. Transformations typically involve temporal resolution harmonisation. 
\end{sloppypar}
The absence of a readily available, freely accessible, and thoroughly quality-checked Italian air quality dataset may present an impediment for researchers. To address this gap, the GRINS\_\allowbreak AQCLIM dataset is specifically designed to provide a transparently processed, well-documented and easily accessible dataset.

\subsection*{Dataset overview}
In Italy, more than 700 land monitoring stations have been working during the last decades. The total number may vary over time as old stations are dismantled and new ones are built. Each station is equipped with various sensors, enabling the measurement of different air pollutants. The GRINS\_\allowbreak AQCLIM dataset comprises data from 744 monitoring stations over 4,017 days, yielding approximately three million rows. Table 1 provides a comprehensive list of the air pollutants included in the dataset, detailing the number of stations that have recorded at least one day of data for each pollutant, alongside their overall minimum, mean, and maximum values, as well as the percentage of missing observations after data cleaning.

\begin{table}[t] \centering 
 \caption{Overview of the air quality (AQ) variables in the GRINS\_\allowbreak AQCLIM after data cleaning and temporal transformation. Min, Mean, and Max are expressed as $\mu g m^{-3}$.} 
  \label{tab_staz_inq} 
\begin{tabular}{@{\extracolsep{5pt}} cccccc} 
\\[-3ex]
\hline \\[-1.8ex] 
 \textbf{Air pollutant} & \textbf{\# stations} & \textbf{Min.} & \textbf{Mean} & \textbf{Max.} & \textbf{NA's (\%)} \\ 
\hline \\[-1.8ex] 
CO & $268$ & $0$ & $0.50$ & $58.20$ & $34$ \\ 
NH$_3$ & $1$ & $0$ & $5.71$ & $17.26$ & $57$ \\ 
NO$_2$ & $710$ & $0$ & $21.71$ & $362.21$ & $27$ \\ 
NO & $336$ & $0$ & $11.44$ & $583.46$ & $65$ \\ 
O$_3$ & $402$ & $0$ & $56.45$ & $756.33$ & $29$ \\ 
PM$_{10}$ & $648$ & $0$ & $24.08$ & $2,575$ & $30$ \\ 
PM$_{2.5}$ & $351$ & $0$ & $15.51$ & $907$ & $35$ \\ 
SO$_2$ & $274$ & $0$ & $3.11$ & $868.42$ & $36$ \\ 
\hline \hline \\[-1.8ex] 
\end{tabular} 
\end{table}

Fig. 1 illustrates the geographical distribution of air quality monitoring stations within the GRINS\_\allowbreak AQCLIM, categorised by station type and area in accordance with European classification standards. In particular, station type refers to the predominant emission sources, while station area indicates the prevalent surrounding land-use context. The GRINS\_\allowbreak AQCLIM dataset includes a total of 744 stations, distributed by type and area as follows. Background stations are the most numerous, with 107 rural, 120 suburban, and 209 urban areas. Industrial stations include 16 rural, 72 suburban, and 15 urban locations. Traffic stations are mainly urban, with 184 stations, while 21 are suburban and 1 is rural.

In addition to air pollutant concentrations, the dataset includes relevant climate information to facilitate modelling applications. It is widely known that atmospheric conditions can heavily impact the air quality and, therefore, climate variables are often considered along with airborne concentrations (see e.g., \cite{rodeschini2024scenario, otto2024spatiotemporal, cameletti2011comparing, bertaccini2012modeling, ignaccolo2014kriging, merk2020estimation}). For instance, temperature and boundary layer height are usually negatively related to air pollutant concentrations. Similarly, we typically observe reduced airborne concentrations during periods with increased precipitation or wind speed.

\begin{figure}
    \centering
    \includegraphics[width=0.7\linewidth]{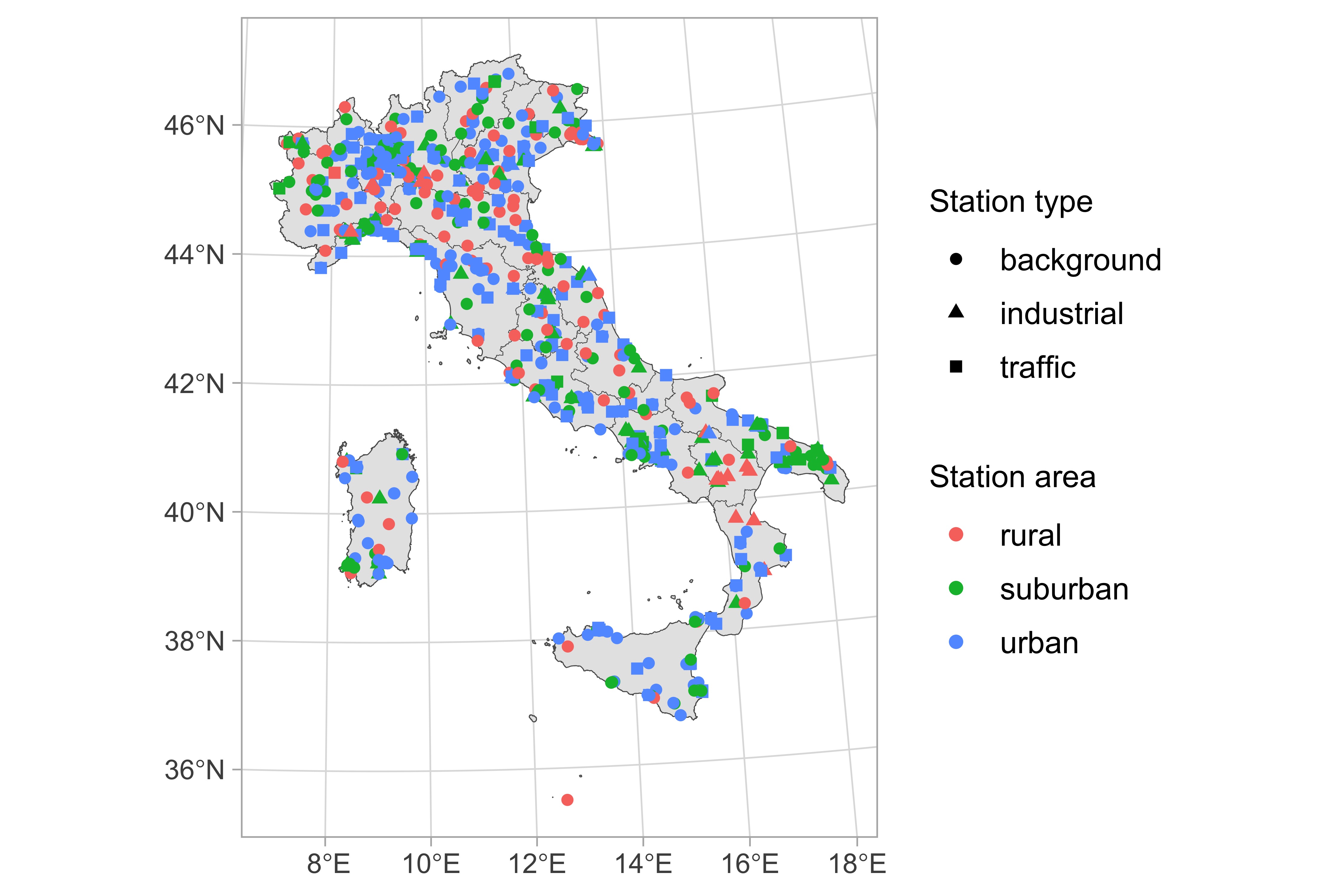}
    \caption{Spatial localisation of the air quality monitoring network in the GRINS\_\allowbreak AQCLIM dataset. Station type refers to the predominant emission sources, while Station area indicates the prevalent surrounding land-use context.}
    \label{fig_staz}
\end{figure}

\section*{Methods}

\subsection*{Input data}
Air quality data are retrieved from the Air Quality Portal - Download Service (\url{https://eeadmz1-downloads-webapp.azurewebsites.net}) managed by the European Environmental Agency (EEA). The portal allows manual or API based download. Air pollutants considered are listed in Table 1. The download of 11 years of data required about 2 hours of time, parallelising the request on 50 workers. It resulted in more than 20k CSV files, each containing a single station for a year and a pollutant. 

Challenges within the above raw files include duplicated measurements (i.e., identical hourly readings) and instances where two distinct raw CSV files present the same station, pollutant, and period. Problematic files exceed 300. The automated procedure illustrated in Fig. 2 was therefore implemented to resolve the duplication issues of such a large volume of raw files.

\begin{sloppypar}
The meteorological data are based on the ERA5 datasets \cite{hersbach2018era5} accessed via the Climate Data Store portal provided by the Copernicus Climate Change Service (\url{https://cds.climate.copernicus.eu/datasets/reanalysis-era5-single-levels?tab=overview}). These datasets contain the numerical model output computed by the European Centre for Medium-Range Weather Forecasts (ECMWF). ERA5 is the fifth generation ECMWF reanalysis of the global climate for the past decades. The reanalysis combines model data with observations from across the world into a globally complete and consistent dataset using the laws of atmospheric science. ERA5 dataset provides hourly estimates for various atmospheric and land-surface quantities with a regular grid scheme at various atmosphere levels on a regular latitude/longitude grid of 0.25°×0.25°. The ERA5Land dataset is obtained by downscaling the ERA5 fields \cite{munoz2021era5} and includes a narrower subset of land-related variables with respect to ERA5, but on a higher spatial resolution (i.e. 0.1°x0.1°). In our study, the ERA5Land dataset was used as the primary source, and ERA5 was employed to fill gaps in areas where ERA5-Land data are unavailable. ERA5Land variables are not available for marine regions, resulting in missing values for some stations along coastlines or on small islands. Therefore, ERA5 and ERA5Land datasets are considered complementary: when ERA5Land data are available, they are always preferable to ERA5 data, while ERA5 fills the gaps for monitoring stations not covered by ERA5Land.
\end{sloppypar}

The climate variables retrieved represent external drivers that typically influence airborne pollutants and are often included as covariates in air-quality modelling studies (e.g. \cite{ignaccolo2014kriging,otto2024spatiotemporal}). The complete set of climate variables is reported in Table 5. In addition to atmospheric conditions, we also retrieved land-surface variables, specifically vegetation cover, distinguishing between high and low vegetation types.

\begin{figure}
    \centering
    \includegraphics[width=0.7\linewidth]{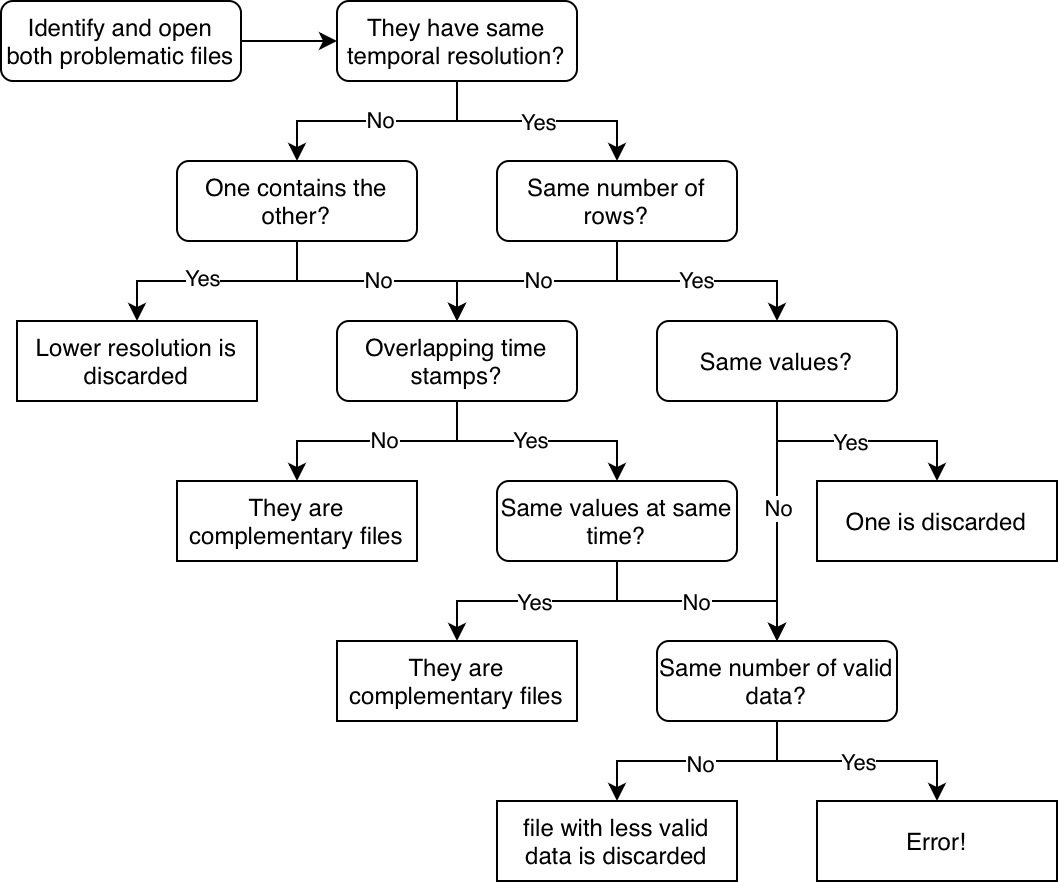}
    \caption{Flow diagram for the automatic fix of duplication issues in raw air quality files from the EEA.}
    \label{fix_CSV}
\end{figure}

\subsection*{Quality check and data cleaning}
\subsubsection*{Anomalies handling}

After automatically resolving conflicts in the raw files, following the procedure explained above, the pollutant concentration values were examined. The EEA provides a validity flag and a verification flag for each measurement. The validity flag indicates the following status: -99 for data invalidated by station maintenance or calibration; -1 for generally invalid data; 1 for valid data; 2 for valid data below the detection limit (BDL); and 3 for valid BDL data replaced by 0.5 times the detection limit. For this work, an initial filter was applied, setting concentrations with a validity flag of -99 or -1 to the missing value.

Despite this preliminary filtering, anomalous concentration values persisted, including extraordinarily high concentrations (e.g., millions of micrograms per cubic metre for particulate matter). An analysis was therefore performed to exclude these values, which likely stemmed from measurement errors, whether technical or operational.

To identify and exclude anomalous values, various thresholds were investigated. Initially, the verification flag was considered. This flag can assume three values: 1 for verified data, 2 for preliminary verified, and 3 for unverified data. The distribution for each pollutant was analysed conditioned on this flag, expecting that anomalous concentrations would primarily fall into the unverified category. However, this was not the case; extreme high values were distributed across categories (for example Table 2 shows that the largest values are obtained by the “verified" case for SO$_2$).

\begin{table}[t]
    \centering
    \caption{Summary Statistics for SO$_2$ concentrations [$\mu g m^{-3}$] by verification flag category.}
    \label{tab:so2_verification_stats}
    \begin{tabular}{l|ccccc}
    \hline
        \textbf{Verification Flag} & \textbf{Min.} & \textbf{Mean} & \textbf{Max.} & \textbf{Sd} & \textbf{Counts (\%)} \\
        \hline
        Verified (1) & 0.0e+00 & 3.5e+00 & 1.4e+06 & 4.4e+02 & 8.4e+01\\
        Preliminary Verified (2) & 1.0e-02 & 3.7e+00 & 3.4e+02 & 3.3e+00 & 2.0e+00\\
        Not Verified (3) & 0.0e+00 & 3.1e+00 & 2.6e+03 & 1.0e+01 & 1.4e+01\\ 
        Overall & 0.0e+00 & 3.5e+00 & 1.4e+06 & 4.0e+02 & 1.0e+02\\
        \hline \hline
    \end{tabular}
\end{table}

To identify appropriate thresholds, multiples of the standard deviation (e.g. 4 $\sigma$) were applied to the overall distribution (2013-2023, all stations) of each pollutant. This z-score method, however, assumes a Gaussian process, a condition often not met by air quality concentration distributions, which frequently exhibit skewness with a long right tail. To mitigate this, the same procedure was applied to logarithmically transformed data. Nevertheless, this alternative approach yielded inconsistent results across pollutants: for instance, the percentage of particulate matter discarded using 6 $\sigma$ on log-transformed data was significantly lower than that for NO$_2$, whereas the opposite occurred when applying the method to untransformed data. Consequently, a standard approach for all pollutants could not be established using this method.

Therefore, the percentiles of the overall distribution, in particular the 99\%, 99.9\%, 99.99\% and 99.999\%, were investigated as possible thresholds. For each percentile, overall statistics were calculated, including the mean, standard deviation, the absolute threshold value, the number of excluded observations, and the count of affected stations.

For visual interpretation, histograms of the excluded observations were generated and inspected for each percentile, with data points coloured according to the recording monitoring station. This aimed to identify potential clusters within the same station that might indicate periods affected by anomalous local conditions. Fig. 3 displays an example of this visual representation, showing that extreme high PM$_{2.5}$ values (over 10,000 $\mu gm^{-3}$) in the entire dataset were recorded at only two stations, located in Alessandria (AL) and Papariano (UD). These values were deemed attributable to technical problems or anomalous local conditions and were subsequently set as missing values.

In conclusion, by jointly considering the tables and histograms (available on the GitHub page, \url{https://github.com/GRINS-Spoke0-WP2/AQ-EEA}), fixed thresholds were established. The thresholds selected, expressed in micrograms per cubic meter, were 100 for CO, 50 for NH$_3$, 1000 for NO, 1000 for NO$_2$, 1000 for O$_3$, 2630 for PM$_{10}$, 980 for PM$_{2.5}$, and 10000 for SO$_2$. These thresholds were applied throughout the analysis to assess air quality levels across the monitoring network: all exceeding concentration values are set as missing.

\begin{figure}
    \centering
    \includegraphics[width=0.9\linewidth]{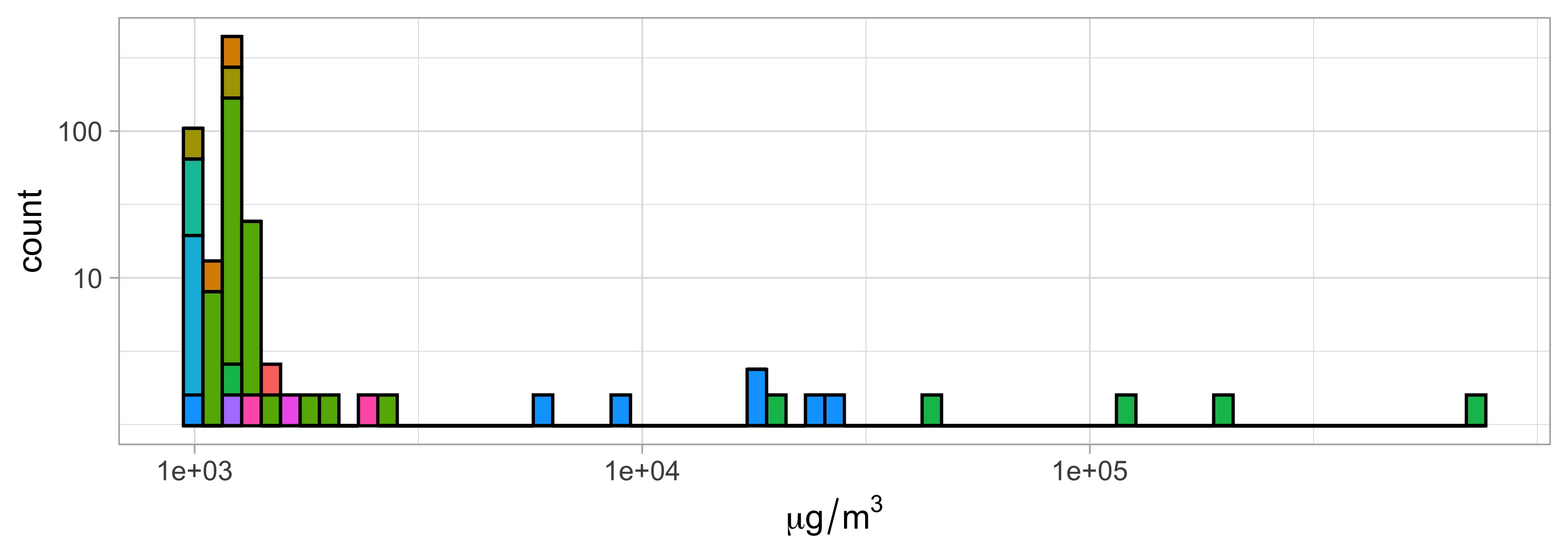}
    \caption{PM${2.5}$ Histograms of observations exceeding the 99.99\% percentiles. Colours according on the monitoring station.}
\end{figure}

\subsubsection*{Monitoring Station Registry}
An important step within the quality check procedure is about the information associated to the monitoring stations. This information encompasses geographical coordinates (longitude and latitude, defining a unique spatial reference point) and the station's classification according to European standards (i.e., by area type and emission source), among other details. During this process, it was identified that three stations shared identical coordinates despite having distinct names and concentration readings. These discrepancies were manually rectified by cross-referencing metadata provided by the regional agency, which also supplied information regarding the sensors installed at each station. Furthermore, 32 stations were found to be duplicated within the EEA station registry, erroneously assigned to two different area categories simultaneously (e.g., classified as both “traffic" and “industrial"). To resolve these inconsistencies, the surrounding area for these stations was manually verified using satellite imagery and subsequently corrected.

\subsection*{Harmonisation of temporal and spatial resolution}
Temporal resolution varies among air quality monitoring stations. In particular, almost all air pollutants are measured hourly, with some exceptions for particulate matter. When data are already on a daily resolution, no further actions are needed. When data are on a hourly basis, before moving toward daily resolution, it was necessary to address hourly missing values to prevent bias in daily summary statistics. Similarly to \cite{fasso2023agrimonia}, this has been done by fitting a local level model \cite{gardner1980algorithm} to hourly data for each station and imputing the missing values by the estimates obtained through the Kalman smoother, as discussed in the Technical validation Section. 

However, daily values were still designated as missing if more than five consecutive hourly data points were missing. To maximise information retention during the conversion from hourly to daily resolution, the minimum, first quartile, mean, median, third quartile, and maximum values were computed. Median, first and third quartiles are calculated using the R function “quantile" from the “stats" package, setting $p$ to $0.5$, $0.25$ and $0.75$ respectively in the function 
\begin{equation}
\label{quartile}
    Q(p)=(np-p-j)x_j + (np+1-p-j)x_{j+1}
\end{equation}
where $x_j$ is the $j$th order statistic of the considered day, $j$ is the integer part of $(np + 1-p)$, $n$ is the sample size, that is, the number of hourly observations for the specific day. 

The hourly climate variables have been converted to the daily resolution using the transformations detailed in Table 3. Moreover, the spatial matching between the air quality monitoring network points and the climate variable grid was performed using the “point-extraction" method. This method consists in associating each monitoring station with the daily climate conditions present within the grid cell to which the station belongs.

\section*{Data Records}

\begin{sloppypar}
Data can be downloaded from the Zenodo page (\url{https://zenodo.org/records/17605148}) or the Amelia platform (\url{https://ameliadp.grins.it/datasetGuest}). Regarding the GRINS\_\allowbreak AQCLIM dataset, the repository contains three main datasets: the dataset on air quality, the information about the land-ground monitoring stations and the missing imputation uncertainty propagated to daily statistics. The file are named GRINS\_\allowbreak AQCLIM\_\allowbreak points\_\allowbreak Italy, Station\_\allowbreak registry\_\allowbreak information and GRINS\_\allowbreak AQCLIM\_\allowbreak imputation\_\allowbreak uncertainty, respectively. All the files are available in CSV and R format.
\end{sloppypar}

\begin{sloppypar}
The GRINS\_\allowbreak AQCLIM\_\allowbreak points\_\allowbreak Italy dataset is divided in two dimensions: Air Quality (AQ) and Climate (CL). Consequently, the column names have two different prefixes, namely AQ\_\allowbreak  and CL\_\allowbreak . The dataset includes the “AirQualityStation" column, which reports the identification code of the station, and the column “time", which identifies the day. There are 48 columns combining the prefix “AQ\_\allowbreak " with summary statistics (i.e. “min", “q1", “mean", “med", “q3", “max") and the pollutant name. For instance, the column name “AQ\_\allowbreak q1\_\allowbreak NO$2$" means the first quartile of the daily distribution of nitrogen dioxide recorded by a particular station on a particular day. Considering the climate part, variables included are listed in Table 3 (more information about climate variables can be found on the European dedicated webpage: \url{https://cds.climate.copernicus.eu/datasets/reanalysis-era5-single-levels?tab=overview}).
\end{sloppypar}

\begin{table}[t]
    \centering
    \caption{Description of daily climate variables included in the GRINS\_\allowbreak AQCLIM.}
    \label{tab:weather_variables}
    \begin{tabular}{llc}
        \toprule
        \textbf{Variable Name} & \textbf{Description} & \textbf{Unit} \\
        \midrule
        CL\_lai\_hv    & Constant fixed value of high vegetation leaf area index & $m^2/m^2$ \\
        CL\_lai\_lv    & Constant fixed value of low vegetation leaf area index & $m^2/m^2$ \\
        CL\_rh         & Mean of relative humidity & $\%$ \\
        CL\_ssr        & Maximum of surface solar radiation & $J/m^2$ \\
        CL\_t2m        & Mean of temperature at 2 meters & $^\circ C$ \\
        CL\_tp         & Total precipitation & $m$ \\
        CL\_winddir    & Mode of wind direction (1=N, 2=E, 3=S, 4=W) & $-$ \\
        CL\_windspeed  & Mean of wind speed & $m/s$ \\
        CL\_blh        & Mean of the height of the atmosphere boundary layer & $m$ \\
        \bottomrule
    \end{tabular}.
\end{table}

The Station\_\allowbreak registry\_\allowbreak information.csv file contains useful information on the monitoring stations, in particular: the “AirQualityStation" field which identifies the name of the station, “Longitude" and “Latitude" are the coordinated in the reference system EPSG 4326 WGS-84, “Altitude" the altitude of the station, “AirQualityStationType" the type of emission sources prevalent, “AirQualityStationArea" the type of area most surrounded it. This information is not changing over time, so they are contained in a separate file to optimise storage.

The estimated imputation uncertainty for daily statistics are provided in the GRINS\_\allowbreak AQCLIM\_\allowbreak imputation\_\allowbreak uncertainty dataset. These estimates are calculated considering both variances and covariances through all the daily summary statistics. In particular, if the minimum or maximum daily values is imputed, the root square of the associated Kalman smoother conditional variance is reported. For quartiles and median, only conditional variances and covariances of the two selected ordered statistics are considered. For the daily average, all the conditional variances and covariances intra-day are considered. More details about missing-data imputation are available in the Technical validation section. We notice that zero values refer to daily averages without missing values at the hourly level, NaN values refer to missing daily averages, and positive uncertainties are related to daily averages with one or more hourly imputed values. In the GRINS\_\allowbreak AQCLIM\_\allowbreak imputation\_\allowbreak uncertainty dataset, only the days when at least one imputation is done are reported to save storage. 

\section*{Technical Validation}
To validate the GRINS\_\allowbreak AQCLIM dataset, comparisons were made with established external data sources. Air quality values were compared with the Agrimonia dataset \cite{fasso2023agrimonia}, a well-known dataset available on Zenodo (\url{https://zenodo.org/records/17605148}) that includes air quality and climate data, among other variables. The Agrimonia dataset was specifically chosen because its air quality measurements are directly obtained from the Lombardy regional agency, rather than from the EEA. Consequently, station names differ, and locations may vary slightly. For the comparison, NO$_2$ measurements with proximate geo-referentation (less than 100m apart in Milan city centre) were selected, assuming they represent the same monitoring station. Fig. 4 illustrates the corresponding two time series during 2021. The time series almost coincide, with the values from the Agrimonia dataset consistently falling within the range defined by the GRINS\_\allowbreak AQCLIM's minimum and maximum values and overlapping with the mean. The root mean squared difference of the two daily averages is 0.8 $\mu g/m^3$.

\begin{figure}
    \centering
    \includegraphics[width=0.9\linewidth]{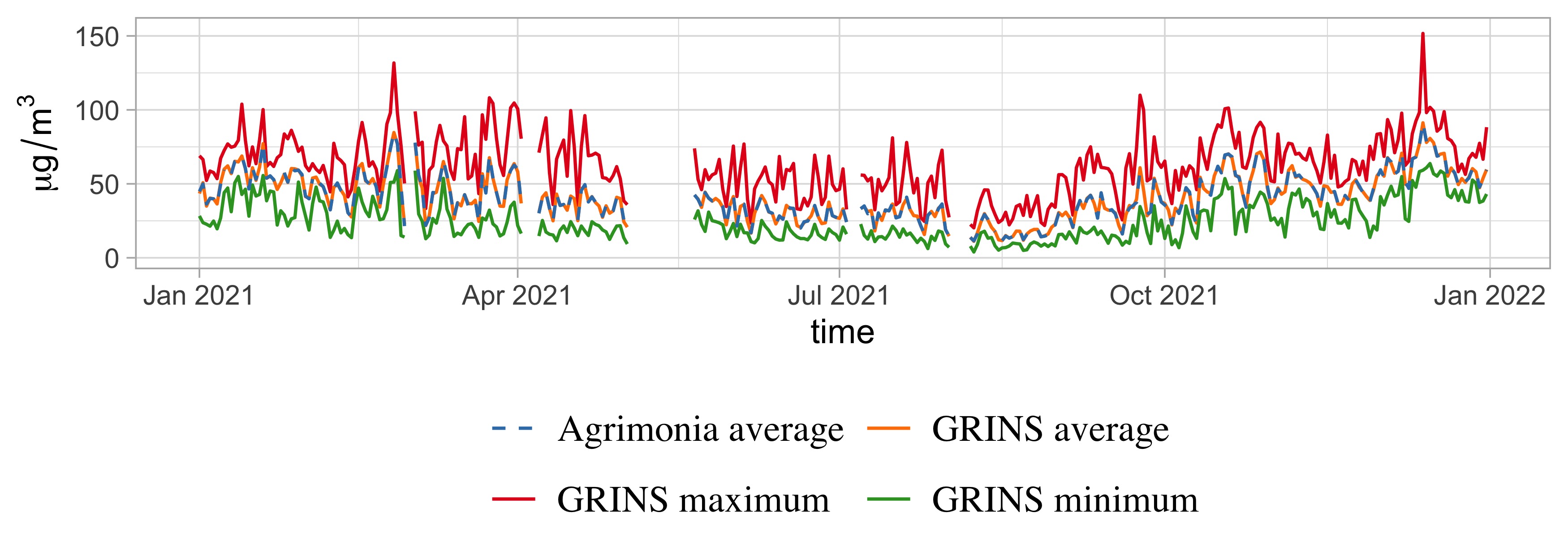}
    \caption{NO$_2$: Comparison of nitrogen dioxide concentrations provided by the Agrimonia and GRINS\_\allowbreak AQCLIM. Data for Milan city centre. RMSE = 0.8 $\mu g/m^3$.}
    \label{fig:tech_val1}
\end{figure}

\begin{sloppypar}
For the climate component of the GRINS\_\allowbreak AQCLIM, which is derived from the ERA5 datasets, validation was performed by comparing temperature with measurements from ground-based monitoring stations (i.e., ARPA). Temperature was selected as a proxy. Ground-based temperature data for the Como area have been manually downloaded from the official website of the regional agency (\url{https://www.arpalombardia.it/temi-ambientali/meteo-e-clima/form-richiesta-dati/}). Fig. 5 presents this comparison, showing the temperature data from the GRINS\_\allowbreak AQCLIM against observations from the ground-based weather station. We notice that ERA5 temperature should be interpreted as an areal average (0.1°$\times$ 0.1°) while ARPA provides local measurements. Hence in the comparison, a change of support issue should be taken into account. Despite of this, the two time series exhibit close values with an RMSE of 1.32 °C.
\end{sloppypar}

\begin{figure}
    \centering
    \includegraphics[width=0.9\linewidth]{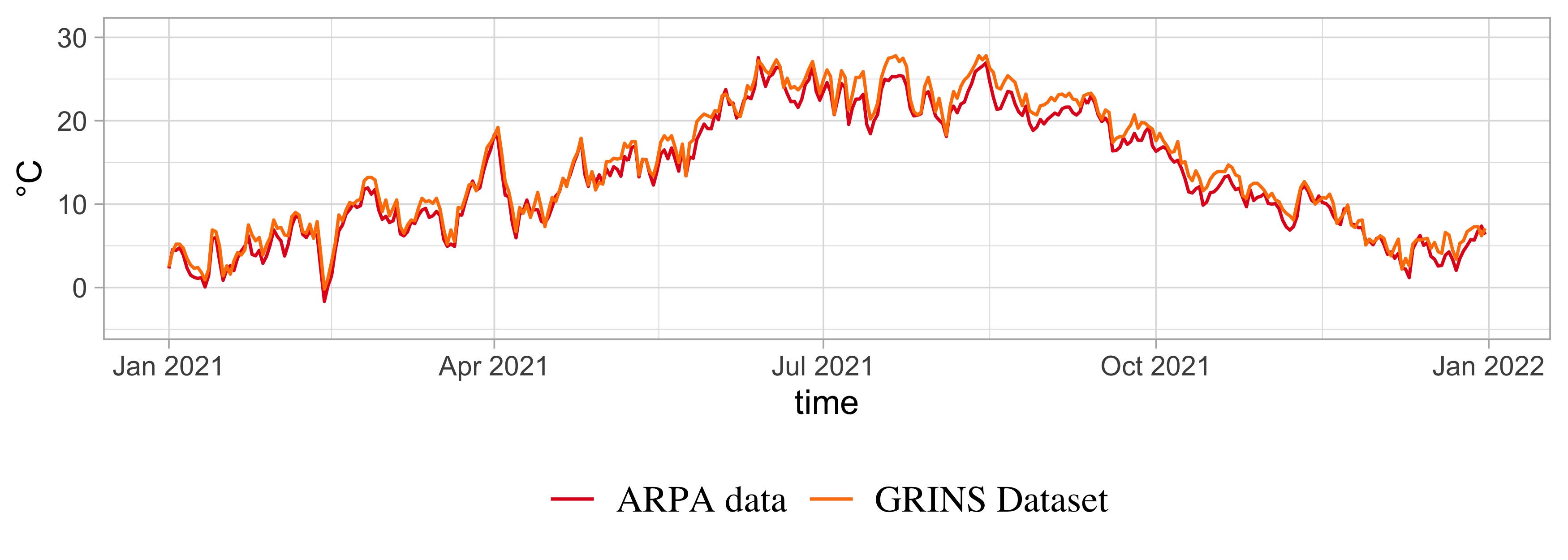}
    \caption{Temperature: Comparison of data provided by regional agency (from ground-based monitoring station) and data contained in the GRINS\_\allowbreak AQCLIM (from ERA5 dataset) for Como city. RMSE = 1.32 °C.}
    \label{fig:tech_val2}
\end{figure}

\subsubsection*{Measurement uncertainty}
Air quality data used here are based on reference measurements. This means they are obtained by well-established measurement instruments and procedures, as detailed in the European Parliament's Directive 2008/50/EC. According to this, the uncertainty (expressed at a 95\% confidence level) should not exceed a certain percentage of the corresponding limit value. The maximum percentage is 15\% for all pollutants except particulate matters, which have a 25\% limit. This must hold for each averaging period considered. Despite this, neither EEA nor local protection agencies provide the uncertainty at the single data level.

ECMWF provides a measure of uncertainty for ERA5, calculated from the ensemble of data assimilations (EDA) system. The EDA accounts primarily for random uncertainties in observations, sea surface temperature and the physical parametrizations of the model. However, systematic model errors are not considered by the EDA. For affordability reasons, the EDA has a lower resolution than ERA5, making it unable to directly describe all of ERA5's uncertainties. Nevertheless, it can be useful for comparing recent days to paste decades or for a given day or season affected by strong weather behaviours, such as areas near tropical cyclones or along storm tracks. Considering the spatial and temporal domain of the dataset, and the absence of strong weather behaviours, we have decided not to include this measure in the dataset. Uncertainty is available on the ERA5 webpage hosted on CDS, under the category ensemble spread (\url{https://cds.climate.copernicus.eu/datasets/reanalysis-era5-single-levels?tab=download}).

\subsubsection*{Missing data imputation and associated uncertainty}
Air quality data from land-ground stations measurements are highly affected by missing values due to a variety of possible causes, either technical or manual. To convert hourly data to the daily resolution, missing hourly values were imputed to avoid bias. For each 11-year hourly time series, all missing values were imputed. However, days containing more than five consecutive hours of missing data were recorded as missing in the daily dataset. To impute the missing hourly data we assume that the air quality concentrations follow a random walk, where the variance of the innovation term is estimated through the L-BFGS-B \cite{liu1989limited} method. L-BFGS-B is a box-constrained and limited-memory version of the well-known Broyden–Fletcher–Goldfarb–Shanno algorithm (BFGS) \cite{fletcher2000practical}.  After the estimation of the variance parameter, the Kalman smoother is run in order to estimate the missing values. An example of this technique is shown in Fig. 6. Clearly, the large variance for the 3rd December results in a missing daily value according to the procedure explained above.

\begin{figure}
    \centering
    \includegraphics[width=0.9\linewidth]{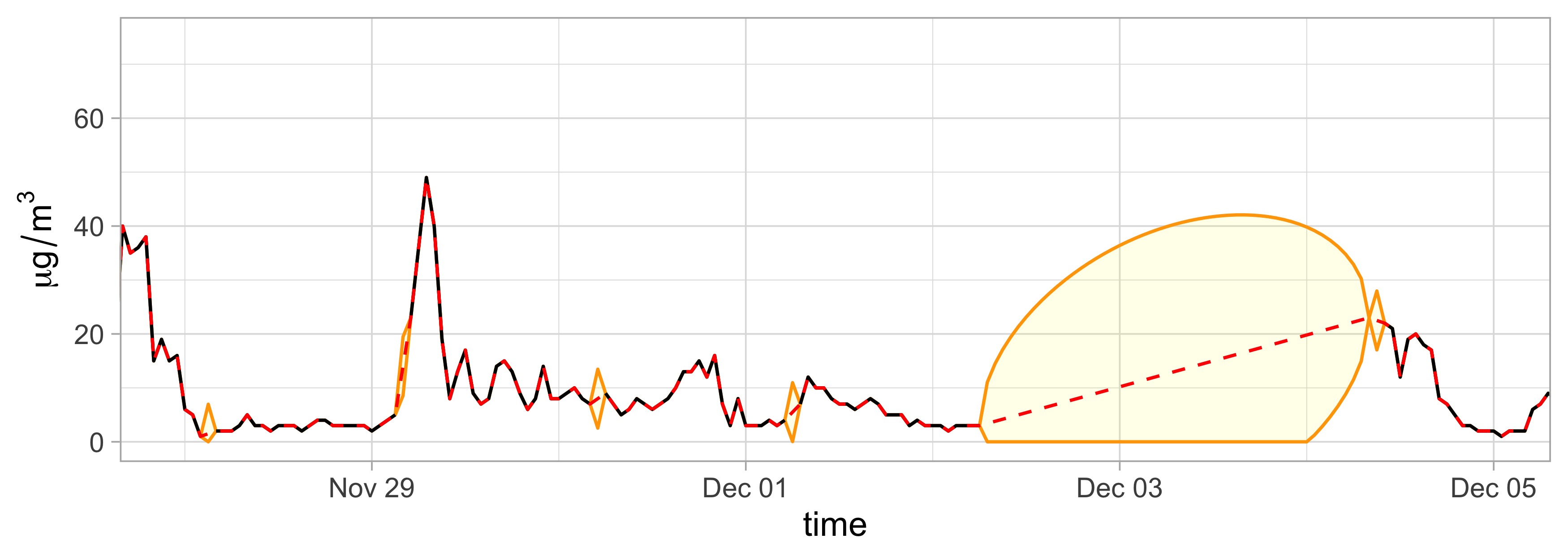}
    \caption{NO$_2$: Example of missing imputation procedure for five days and one station measuring NO$_2$. Black line is observed data, red dotted line is the imputed values, orange bands represents the standard deviation.}
    \label{fig:imputation}
\end{figure}

Along with daily values we provide uncertainty quantification. The variance of the daily mean depends on the conditional variances and covariances of the imputed values. In particular, considering the day $d$ and its 24 hours, the variance of the daily average $\bar{y}_d$ is calculated as:

\begin{equation}
Var(\bar{y}_d)= \frac{1}{24^2}\left( 
\sum_{i\in d} Var(y_i) 
+ 
2\sum_{\substack{i,j \in d \\ i \neq j}} Cov(y_i,y_j) 
\right)
\end{equation}

where $y_i, \ i \in d,$ are the hourly concentrations of day $d$. In our case, the variances of the observed data are set to 0 while the variances and the covariances of imputed values are derived from the Kalman smoother (see e.g. \cite{durbin2012time} for more details). In particular, for $n$ data and $i = \{1,\ldots,n\}$, assume $P_i$ is the conditional variance of the Kalman filter at time $i$ (i.e., $Var(y_i|Y_{i-1})$), and $V_i$ the conditional variance of the Kalman smoother (i.e., $Var(y_i|Y_{n})$), where $Y_j=(y_1,\ldots,y_j)^T$. The variances of imputed values are $V_i$ while the covariances are calculated as:
\begin{equation}
Cov(y_i,y_j) = P_i(L_i \cdots L_{i-j})(1-N_{j-1}P_j) \quad \text{for} \quad i,j \in \tau \quad \text{and} \quad i>j
\end{equation}
where $\tau \subseteq \{1,\ldots,n\}$ are the time stamps related to missing observations of day $d$, $N_{j-1} = P_j^{-1} - L_j^2 N_{j}$ is obtained recursively setting $N_n = 0$, and $L_j$ is set to 1 for missing values (i.e. $j \in \tau$) and 0 otherwise. It is evident that covariances are different from 0 only for consecutively missing data.

The same approach is used to calculate the uncertainty of the considered daily quantiles, namely the 1st, 2nd and 3rd quartiles. Here variances and covariances are calculated according to Formula 1. If the daily minimum or maximum is an imputed value, the uncertainty associated is the related smoothed variance. 

\section*{Data availability}

\begin{sloppypar}
Data can be downloaded from the Zenodo page (\url{https://zenodo.org/records/17605148}) and they will be accessible through the Amelia platform (\url{https://ameliadp.grins.it/datasetGuest}). The repository contains three main datasets: the dataset on air quality, the information about the land-ground monitoring stations and the missing imputation uncertainty propagated to daily statistics. The file are named GRINS\_\allowbreak AQCLIM\_\allowbreak points\_\allowbreak Italy, Station\_\allowbreak \allowbreak registry\_\allowbreak \allowbreak information and GRINS\_\allowbreak AQCLIM\_\allowbreak imputation\_\allowbreak uncertainty, respectively. All the files are available in CSV and R format.
\end{sloppypar}

The GRINS\_\allowbreak AQCLIM\_\allowbreak points\_\allowbreak Italy dataset is divided in two dimensions: Air Quality (AQ) and Climate (CL). The names of the columns have two different suffixes, following this division (i.e. AQ\_\allowbreak  and CL\_\allowbreak ). The Station\_\allowbreak registry\_\allowbreak information dataset contains useful information about the monitoring stations (e.g. georeferentiation). The GRINS\_\allowbreak AQCLIM\_\allowbreak imputation\_\allowbreak uncertainty dataset contains estimated imputation uncertainty for daily statistics.

\section*{Code Availability}
All the code used to generate the dataset is available at the GitHub page of the GRINS project (\url{https://github.com/GRINS-Spoke0-WP2/AQ-EEA} for air quality, and \url{https://github.com/GRINS-Spoke0-WP2/WE-C3S} for climate), online and freely consultable. Moreover, on the GitHub page there is extensive documentation that explains the R scripts. The software version used is R 24.3.0.

\section*{Acknowledgements, Author Contributions \& Competing Interests}

\subsection*{Acknowledgements}
\begin{sloppypar}
\textbf{Funder}: Project funded under the National Recovery and Resilience Plan (NRRP), Mission 4 Component 2 Investment 1.3 - Call for tender No. 341 of 15/03/2022 of Italian Ministry of University and Research funded by the European Union – NextGenerationEU \\
\textbf{Award Number}: PE\_\allowbreak 00000018, Concession Decree No. 1558 of 11/10/2022 adopted by the Italian Ministry of University and Research, CUP F83C22001720001, GROWING RESILIENT INCLUSIVE AND SUSTAINABLE - GRINS \\
\end{sloppypar}

\subsection*{Author contributions}
AFM drafted the manuscript, performed data acquisition and data analysis, prepared the figures and tables. AF conceived and supervised the study, reviewed the manuscript. JR reviewed the manuscript. All authors read and approved the final manuscript.

\subsection*{Competing interests}
The author(s) declare no competing interests


\begin{thebibliography}{99}

\bibitem{eu_quality}
Targa, J., Colina, M., Banylus, L., González Ortiz, A., \& Soares, J. Status report of air quality in Europe for year 2024, using validated and up-to-date data (ETC-HE Report 2025/1). European Topic Centre on Human health and the environment (2024).

\bibitem{otto2024spatiotemporal}
Otto, P., Fusta Moro, A., Rodeschini, J., Shaboviq, Q., Ignaccolo, R., Golini, N., Cameletti, M., Maranzano, P., Finazzi, F., \& Fassò, A. Spatiotemporal modelling of PM${2.5}$ concentrations in Lombardy (Italy): a comparative study. \emph{Environmental and Ecological Statistics} \textbf{31}, 245–272 (2024).

\bibitem{capobussi2016air}
Capobussi, M., Tettamanti, R., Marcolin, L., Piovesan, L., Bronzin, S., Gattoni, M. E., Polloni, I., Sabatino, G., Tersalvi, C. A., Auxilia, F., et al. Air pollution impact on pregnancy outcomes in Como, Italy. \emph{J. Occup. Environ. Med.} \textbf{58}, 47–52 (2016).

\bibitem{coker2020effects}
Coker, E. S., Cavalli, L., Fabrizi, E., Guastella, G., Lippo, E., Parisi, M. L., Pontarollo, N., Rizzati, M., Varacca, A., \& Vergalli, S. The effects of air pollution on COVID-19 related mortality in northern Italy. \emph{Environmental and Resource Economics} \textbf{76}, 611–634 (2020).


\bibitem{gatti2020machine}
Gatti, R. C., Velichevskaya, A., Tateo, A., Amoroso, N., \& Monaco, A. Machine learning reveals that prolonged exposure to air pollution is associated with SARS-CoV-2 mortality and infectivity in Italy. \emph{Environmental Pollution} \textbf{267}, 115471 (2020).

\bibitem{bozzani2024spatio}
Bozzani, A., Cutti, S., Di Marzo, L., Gabriele, R., \& Sterpetti, A. V. Spatio-temporal correlation between admissions for ruptured abdominal aortic aneurysms and levels of atmospheric pollution in Italy. \emph{Current Problems in Cardiology} \textbf{49}, 102249 (2024).

\bibitem{conti2018association}
Conti, S., Harari, S., Caminati, A., Zanobetti, A., Schwartz, J. D., Bertazzi, P. A., Cesana, G., \& Madotto, F. The association between air pollution and the incidence of idiopathic pulmonary fibrosis in Northern Italy. \emph{European Respiratory Journal} \textbf{51}, 1 (2018).

\bibitem{trentalange2025association}
Trentalange, A., Badaloni, C., Porta, D., Michelozzi, P., \& Renzi, M. Association between air quality and neurodegenerative diseases in River Sacco Valley: A retrospective cohort study in Latium, central Italy. \emph{Int. J. Hygiene Environ. Health} \textbf{267}, 114578 (2025).

\bibitem{bottalico2016air}
Bottalico, F., Chirici, G., Giannetti, F., De Marco, A., Nocentini, S., Paoletti, E., Salbitano, F., Sanesi, G., Serenelli, C., \& Travaglini, D. Air pollution removal by green infrastructures and urban forests in the city of Florence. \emph{Agriculture and Agricultural Science Procedia} \textbf{8}, 243–251 (2016).

\bibitem{manes2016regulating}
Manes, F., Marando, F., Capotorti, G., Blasi, C., Salvatori, E., Fusaro, L., Ciancarella, L., Mircea, M., Marchetti, M., Chirici, G., et al. Regulating ecosystem services of forests in ten Italian metropolitan cities: air quality improvement by PM10 and O3 removal. \emph{Ecological Indicators} \textbf{67}, 425–440 (2016).

\bibitem{sebastiani2021modeling}
Sebastiani, A., Buonocore, E., Franzese, P. P., Riccio, A., Chianese, E., Nardella, L. \& Manes, F. Modeling air quality regulation by green infrastructure in a Mediterranean coastal urban area: The removal of PM10 in the Metropolitan City of Naples (Italy). \textit{Ecol. Model.} \textbf{440}, 109383 (2021).

\bibitem{rodeschini2024scenario}
Rodeschini, J., Fassò, A., Finazzi, F., \& Fusta Moro, A. Scenario analysis of livestock-related PM2.5 pollution based on a new heteroskedastic spatiotemporal model. \emph{Socio-Economic Planning Sciences} \textbf{96}, 102053 (2024).

\bibitem{fasso2023agrimonia}
Fassò, A., Rodeschini, J., Fusta Moro, A., Shaboviq, Q., Maranzano, P., Cameletti, M., Finazzi, F., Golini, N., Ignaccolo, R., \& Otto, P. Agrimonia: a dataset on livestock, meteorology and air quality in the Lombardy region, Italy. \emph{Scientific Data} \textbf{10}, 143 (2023).

\bibitem{marecal2015regional}
Marécal, V., Peuch, V.-H., Andersson, C., Andersson, S., Arteta, J., Beekmann, M., Benedictow, A., Bergström, R., Bessagnet, B., Cansado, A., et al. A regional air quality forecasting system over Europe: the MACC-II daily ensemble production. \emph{Geoscientific Model Development} \textbf{8}, 2777–2813 (2015).

\bibitem{CAMS_EQC_IRA_2022}
Copernicus Atmosphere Monitoring Service (CAMS). Evaluation and Quality Control (EQC) for the CAMS Regional Production System - Interim Report 2021. \url{https://atmosphere.copernicus.eu/sites/default/files/custom-uploads/EQC-regional/IRA/CAMS283_2022SC1_D83.2.1.1-2021_202208_EQC_IRA2021_v1.pdf (2022)}.

\bibitem{wei2021chinahighpm10}
Wei, J., Li, Z., Xue, W., Sun, L., Fan, T., Liu, L., Su, T., \& Cribb, M. The ChinaHighPM10 dataset: generation, validation, and spatiotemporal variations from 2015 to 2019 across China. \emph{Environment International} \textbf{146}, 106290 (2021).

\bibitem{wei2023ground}
Wei, J., Li, Z., Wang, J., Li, C., Gupta, P. \& Cribb, M. Ground-level gaseous pollutants (NO$_2$, SO$_2$, and CO) in China: daily seamless mapping and spatiotemporal variations. \textit{Atmos. Chem. Phys.} \textbf{23}, 1511--1532 (2023).

\bibitem{fioravanti2022spatiotemporal}
Fioravanti, G., Cameletti, M., Martino, S., Cattani, G. \& Pisoni, E. A spatiotemporal analysis of NO$_2$ concentrations during the Italian 2020 COVID-19 lockdown. \textit{Environmetrics} \textbf{33}, e2723 (2022).

\bibitem{cameletti2011comparing}
Cameletti, M., Ignaccolo, R. \& Bande, S. Comparing spatio-temporal models for particulate matter in Piemonte. \textit{Environmetrics} \textbf{22}, 985--996 (2011).

\bibitem{bertaccini2012modeling}
Bertaccini, P., Dukic, V. \& Ignaccolo, R. Modeling the short-term effect of traffic and meteorology on air pollution in Turin with generalized additive models. \textit{Adv. Meteorol.} \textbf{2012}, 609328 (2012).

\bibitem{ignaccolo2014kriging}
Ignaccolo, R., Mateu, J. \& Giraldo, R. Kriging with external drift for functional data for air quality monitoring. \textit{Stoch. Environ. Res. Risk Assess.} \textbf{28}, 1171--1186 (2014).

\bibitem{merk2020estimation}
Merk, M. S. \& Otto, P. Estimation of anisotropic, time-varying spatial spillovers of fine particulate matter due to wind direction. \textit{Geogr. Anal.} \textbf{52}, 254--277 (2020).

\bibitem{hersbach2018era5}
Hersbach, H., Bell, B., Berrisford, P., Biavati, G., Horányi, A., Muñoz Sabater, J., Nicolas, J., Peubey, C., Radu, R., Rozum, I., et al. ERA5 hourly data on single levels from 1979 to present. \emph{Copernicus Climate Change Service (C3S) Climate Data Store (CDS)} \textbf{10}, 10.24381 (2018).

\bibitem{munoz2021era5}
Muñoz-Sabater, J., Dutra, E., Agustí-Panareda, A., Albergel, C., Arduini, G., Balsamo, G., Boussetta, S., Choulga, M., Harrigan, S., Hersbach, H., et al. ERA5-Land: A state-of-the-art global reanalysis dataset for land applications. \textit{Earth Syst. Sci. Data} \textbf{13}, 4349--4383 (2021).

\bibitem{gardner1980algorithm}
Gardner, G., Harvey, A. C., \& Phillips, G. D. A. Algorithm AS 154: An algorithm for exact maximum likelihood estimation of autoregressive-moving average models by means of Kalman filtering. \emph{J. R. Stat. Soc. Series C (Applied Statistics)} \textbf{29}, 311–322 (1980).

\bibitem{liu1989limited}
Liu, D. C. \& Nocedal, J. On the limited memory BFGS method for large scale optimization. \textit{Math. Program.} \textbf{45}, 503--528 (1989).

\bibitem{fletcher2000practical}
Fletcher, R. \textit{Practical Methods of Optimization}. John Wiley \& Sons, Chichester, UK (2000).

\bibitem{durbin2012time}
Durbin, J. \& Koopman, S. J. \textit{Time Series Analysis by State Space Methods}. Oxford University Press, Oxford, UK (2012).

\end{thebibliography}
\end{document}